\documentclass[%
prd,
reprint,
superscriptaddress,
preprintnumbers,
nofootinbib,
amsmath,amssymb,
aps,
]{revtex4-1}
\usepackage{hyperref}
\usepackage{graphicx}

\newcommand{\eV}{\text{\,eV}}
\newcommand{\keV}{\text{\,keV}}

\newcommand{\be}{\begin{equation}}
\newcommand{\ee}{\end{equation}}
\newcommand{\ba}{\begin{eqnarray}}
\newcommand{\ea}{\end{eqnarray}}

\renewcommand{\l}{\left(}
\renewcommand{\r}{\right)}
\newcommand{\la}{\langle}
\newcommand{\ra}{\rangle}
\newcommand{\e}{\mathrm{e}}

\newcommand{\DDM}{\text{DDM}}
\newcommand{\DM}{\text{DM}}

\newcommand{\vev}[1]{\langle #1\rangle}

\begin{document}

\preprint{INR-TH/2014-030,CERN-PH-TH-2014-246}

\title{On the applicability of approximations used in calculation of
  spectrum of Dark Matter particles produced in particle decays}

\author{Fedor Bezrukov}
\email{fedor.bezrukov@uconn.edu} 
\affiliation{CERN, CH-1211 Gen\`eve 23, Switzerland}
\affiliation{Physics Department, University of Connecticut, Storrs, CT
  06269-3046, USA}
\affiliation{RIKEN-BNL Research Center, Brookhaven National
  Laboratory, Upton, NY 11973, USA}

\author{Dmitry Gorbunov}
\email{gorby@ms2.inr.ac.ru}
\affiliation{Institute for Nuclear Research of the Russian Academy of
  Sciences, Moscow 117312, Russia}
\affiliation{Moscow Institute of Physics and Technology, 
  Dolgoprudny 141700, Russia}


\begin{abstract}
  For the Warm Dark Matter (WDM) candidates the momentum distribution
  of particles becomes important, since it can be probed with
  observations of Lyman-$\alpha$ forest structures and confronted with
  coarse grained phase space density in galaxy clusters.  We recall
  the calculation \cite{Kaplinghat:2005sy} of the spectrum in case of
  dark matter non-thermal production in decays of heavy particles
  emphasizing on the inherent applicability conditions, which are
  rather restrictive and sometimes ignored in literature.  Cold part
  of the spectrum requires special care when WDM is considered.
\end{abstract}

\maketitle

\section{Introduction}

One of the major puzzles in physics as we know it at present---Dark
Matter (DM) phenomenon---requires new massive electrically neutral
collisionless particles stable at cosmological
time-scale~\cite{Trimble:1987ee}.  They must be produced in the early
Universe before the plasma temperature $T$ drops below 1\eV, since
later cosmological stages definitely need the
DM~\cite{Hinshaw:2012aka}.

While the Universe expansion schedule is sensitive only to the total
energy density associated with the new non-relativistic particles (and
hence to their number density at a given particle mass), the evolution
of spatial \emph{inhomogeneities} of matter is also sensitive to the
velocity distribution of the DM particles.  Indeed, free streaming of
the DM particles smooths out all the inhomogeneities smaller than the
so-called free-streaming length $l_{\text{f.s.}}$.  The latter is the
typical distance travelled by a DM particle, which is of order
$l_{\text{f.s.}}\sim v\times l_H$, where $v$ is the DM average
velocity and $l_H$ is the Hubble horizon size at a given time.  In
order for successful generation of the smallest observed primordial
structures---dwarf galaxies---one needs $v\lesssim 10^{-3}$ at the
epoch of radiation-matter energy density equality, $T\sim 1$\eV.  This
requirement defines the border line between faster and slower
candidates named as Hot and Cold DM .

The candidates \emph{right at the border} are called Warm DM, and the
question about velocity distribution is especially relevant for them.
In fact, the Hot DM is disfavoured by structure formation and may be
only a small fraction of DM (precise amount depends on the velocity
distribution).  On the other extreme, the Cold DM candidates, like
Weakly Interacting Massive Particles, are typically very slow at
equality, and allow for formation of structures much smaller (and
lighter) than the dwarf galaxies.  These structures are expected to be
starless and empty of baryons after reionization epoch and (partially)
destroyed during subsequent formation of heavier structures.  Yet if
some of them remained, searches for gravitational lensing events in
galaxies may (in principle) detect them and determine the structure
abundance (the DM velocity distribution defines the size of smallest
structures).  Therefore, both Hot and Cold DM components suggest
potential observables sensitive to the velocity distribution.
However, this is the intermediate case of Warm DM, where the
corresponding observables provide the most non-trivial constraints on
the DM models, and they have been actively exploited in the
literature.

The most promising observable for this task is the small structures in
the Lyman-$\alpha$ forest~\cite{Viel:2005qj,Viel:2007mv}.  Their
studies have already allowed to rule out some WDM models, e.g.\ (keV
scale) sterile neutrino DM \cite{Viel:2006kd} produced non-thermally
by non-resonant oscillations of active neutrino in primordial plasma
\cite{Dodelson:1993je}.  However, many other candidates are still
valid (e.g.\ light gravitino \cite{Gorbunov:2008ui}, axino
\cite{Choi:2013lwa}, etc.), which are mostly non-thermally produced,
see \cite{Baer:2014eja} for a review.  Moreover, even in the case of
sterile neutrino DM various other production mechanisms were proposed,
such as resonant production \cite{Shi:1998km}, thermal production with
subsequent dilution \cite{Bezrukov:2009th}, production in decays of
scalar particles
\cite{Kaplinghat:2005sy,Shaposhnikov:2006xi,Kusenko:2006rh,Petraki:2007gq},
all providing some ways to evade the present Lyman-$\alpha$
constraints \cite{Boyarsky:2008xj,Viel:2013fqw}.

To use the observations of Lyman-$\alpha$ forest structure in a
particular model one must know the velocity distribution of the DM
particles.  In this note we focus our attention on the DM produced in
the early Universe in decays of some particle, which we, following
\cite{Kaplinghat:2005sy}, denote as DDM.  Several regimes are
possible, corresponding to the DDM particle being in or out of thermal
equilibrium.  The case of DDM in thermal equilibrium due to
annihilations channel in the SM particles, while also having a small
decay branching ratio into the DM, is analysed in
\cite{Shaposhnikov:2006xi,Kusenko:2006rh}.  The production happens
mostly at temperature of about the DDM mass, $T\sim M$, leading to the
distribution with average momentum slightly below the thermal one.
After production the spectrum can be cooled further due to the
decrease of degrees of freedom of the relativistic plasma in the
expanding Universe \cite{Kusenko:2006rh}.  This mechanism leads to the
lower bound on the DM mass in the model, $m_\DM>7.8\keV$ where we used
the recent Lyman-$\alpha$ analysis from \cite{Viel:2013fqw}.  Another
situation corresponds to the case of DDM decaying while being
\emph{out of thermal equilibrium,} \cite{Kaplinghat:2005sy} (in
particular, this may correspond to the DDM itself produced in a
non-thermal way).  It was argued, that for sufficiently long living
DDM, the majority of its decays happen when it is significantly
non-relativistic, leading to a peculiar momentum distribution, that
can be strongly shifted towards low momenta.  In this note we show,
that the approximation of non-relativistic decay is actually valid
only for the high energy part of the DM spectrum, while the low energy
part is produced at earlier stages, when DDM particles still have
non-negligible velocities.

The results of the paper shows, that for the proper description of the
cold parts of DM it is important to analyse exactly the decays of the
DDM at early times, when it is close to being relativistic (no matter
whether it is in or out of thermal equilibrium).  The detailed
analysis of several possibilities of this type is present in
\cite{Kawasaki:1992kg,Shaposhnikov:2006xi,Kusenko:2006rh,Petraki:2007gq,Merle:2015oja}.

In section \ref{sec:gf} we introduce the generic formalism, and review
the non-relativistic approximation used to obtain the DM spectrum in
section \ref{sec:anal}.  Regions of applicability of this
approximation and comparison with exact numerical results are
presented in section \ref{sec:concl}.


\section{General formalism}
\label{sec:gf}

In the model we have two sets of particles---the decaying DDM of mass
$M$ and the dark matter DM, which is stable with mass $m_{\text{DM}}$.
DM is produced in two body decay of the initial DDM particle.
Distribution of the particles over momentum $f(p)$ are normalized to
the physical particle number density $n$ in the expanding Universe
with scale factor $a$ as
\begin{equation}
  \label{normalization}
  n = \!\!\int\! \frac{d^3p}{(2\pi)^3} f(pa)
  = \!\!\int\! \frac{d^3k}{(2\pi)^3a^3} f(k) 
  = \!\!\int\limits_0^\infty\! \frac{k^2dk}{2\pi^2a^3} f(k),
\end{equation}
where $p$ and $k\equiv pa$ are physical and conformal 3-momenta,
respectively.  We allow to be sloppy in writing the conformal or
physical momentum as an argument to $f$, as far as they can always be
mapped to another.  One should just be careful in the solution of the
kinetic equations, where the conformal momentum must be always used.
Also we often drop the explicit time dependence where it does not lead
to ambiguities.  We use conformal $\eta$ and physical $t$ times
(related by $dt=ad\eta$) interchangeably.  The normalization used
corresponds to $f(k)$ remaining constant in time in the absence of
interactions, and (physical) number density decreasing as
$n\propto 1/a^3$.

The kinetic equation for the DDM evolution is (c.f.\ eq.~(1) of
\cite{Kaplinghat:2005sy})
\begin{equation}
  \label{eqDDM}
  \frac{df_\DDM(k_\DDM,\eta)}{d\eta} =
  -\frac{aM}{\tau E_\DDM} f_\DDM(k_\DDM,\eta),
\end{equation}
where $\tau$ is the DDM lifetime.  The equation for the DM is
\begin{multline}
  \label{DMeq}
  \frac{df_\DM(k_\DM,\eta)}{d\eta} =
  \frac{aM^2}{\tau E_\DM p_\DM p_\text{CM}}
  \int_{E_1}^{E_2} f_\DDM(a p) dE \\
  \stackrel{m_\DM\to0}{=}\;
  \frac{a^32M}{\tau k_\DM^2}
  \int_{p_\DM+\frac{M^2}{4p_\DM}}^\infty f_\DDM(a p) dE,
\end{multline}
where $p_\DM\equiv k_\DM/a$, $p\equiv\sqrt{E^2-M^2}$, and
$p_{\text{CM}}$ is the DM momentum in the centre of mass frame of DDM,
$E_{1,2}$ are given by (\ref{E12}).  There is overall coefficient 2 as
compared to (2) of \cite{Kaplinghat:2005sy}, which assumed that only
one of the two-body decay products is the DM.  In the present note we
assume that both DDM decay products are DM.


\section{Analytical solutions to the equations in case of
  radiation domination and constant $g_*$}
\label{sec:anal}

Solution to eq.~\eqref{eqDDM} at the radiation dominated stage with
scale factor $a\equiv c\eta$ is (c.f.\ Ref.~\cite{Petraki:2007gq})
\begin{multline}
  \label{sol-2}
  f_{\text{DDM}}(k,\eta)=f_{\text{DDM}}(k,\eta_i)
  \l
  \frac{\eta+\sqrt{\eta^2+\frac{k^2}{M^2c^2}}}
  {\eta_i+\sqrt{\eta_i^2+\frac{k^2}{M^2c^2}}}
  \r^{\frac{k^2}{2\tau c M^2}} \\
  \times
  \e^{-\frac{c}{2\tau}\l \eta\sqrt{\eta^2+\frac{k^2}{M^2c^2}}- 
    \eta_i\sqrt{\eta_i^2+\frac{k^2}{M^2c^2}}\r}.
\end{multline}
Hereafter the subscript $i$ refers to the moment when the DDM
particles freeze out or appear in the Universe through another
mechanism, leading to some fixed spectrum $f_{\text{DDM}}(k,\eta_i)$.
The analytical solution \eqref{sol-2} assumes constant number of
relativistic degrees of freedom in plasma from the DDM production
$\eta_i$ till its decay.  The solution \eqref{sol-2} can be rewritten
through the physical momenta $p\equiv k/a$ and the Hubble parameter
given by
\begin{equation}
  \label{Hubble}
  H\equiv\frac{da/d\eta}{a^2}=\frac{1}{c\eta^2}.
\end{equation}

In the limit of very non-relativistic particles one obtains
approximately from \eqref{sol-2}
\begin{equation}
  \label{non-rel}
  f_{\text{DDM}}(k,\eta)=f_{\text{DDM}}(k,\eta_i)\times
  \e^{-\frac{1}{2\tau}\l \frac{1}{H}-\frac{1}{H_i}\r},
\end{equation}
and at the next-to-leading order both for the exponent (which remains
the same) and the prefactor
\begin{multline}
  \label{non-rel-correction}
  f_{\text{DDM}}(k,\eta)=f_{\text{DDM}}(k,\eta_i)\\
  \times
  \l 1+\frac{k^2}{a^2M^2}\frac{1}{4\tau H}\log\frac{H_i}{H}\r
  \e^{-\frac{1}{2\tau}\l \frac{1}{H}-\frac{1}{H_i}\r}.
\end{multline}

To solve \eqref{DMeq} one must evaluate the upper $E_2$ and lower
$E_1$ limits of the integration in the r.h.s.  One gets approximately
(in the relativistic limit, $p_\text{DM}\gg m_\DM$)
\begin{align}
  \label{E12}
  E_2 &= p_{\text{DM}}\frac{M^2}{m_{\DM}^2} - p_{\text{DM}} +
  \frac{M^2}{4p_{\text{DM}}}\to\infty
  ,\\\notag
  E_1 &= p_{\text{DM}} +\frac{M^2}{4p_{\text{DM}}}. 
\end{align}

In the non-relativistic limit \eqref{non-rel}, when all decaying
particles DDM are (almost) at rest, it is reasonable to assume that
their distribution function is
\begin{equation}
  \label{condensate}
  f_{\text{DDM}}(k,\eta) = F \frac{2\pi^2\delta(k)}{k^2},
\end{equation}
where the normalization is fixed by (\ref{normalization}), and
\begin{equation}
  \label{Fnrdecay}
  F(\eta) = F_i \times\e^{
    -\frac{1}{2\tau}\left(\frac{1}{H}-\frac{1}{H_i}\right)}
  \equiv \widetilde F_i \times \e^{-\frac{1}{2\tau H}}.
\end{equation}
To avoid singularities at $E=E_1$ in \eqref{DMeq}, it is convenient to
regularize \eqref{condensate} as
\begin{equation}
  \label{fnrreg}
  f_{\text{DDM}}(k) = F \frac{2\pi^2\delta(k-\kappa)}{k^2}.
\end{equation}
This regularization has physical meaning, as it assumes that the DDM
particles are not exactly at rest, but move with some small conformal
momentum $\kappa$.  Note that the normalization \eqref{Fnrdecay}
means, that at the moment $\eta_i$ of DDM freeze-out its concentration
is given by $n_{\text{DDM}}(\eta_i) = F_i/a_i^3$.  In this
approximation the collision integral in \eqref{DMeq} can be taken
easily
\begin{equation}
  \label{NrCi}
  \int\limits_{p_\DM+\frac{M^2}{4p_\DM}}^\infty f_\DDM(a p)\, dE
  = \frac{2\pi^2 F}{a^3 M} \,\delta\!\left(p_\DM-\frac{M}{2}\right).
\end{equation}

Using eq.~\eqref{NrCi} eq.~\eqref{DMeq} can be reduced to
\begin{equation}
  \label{regeq}
  \frac{df_{\text{DM}}(k)}{d\eta}
  \!=\!
  \frac{4\pi^2 F}{\tau k^2} \delta\!\left(\!\frac{k}{a}-\frac{M}{2}\!\right)
  \!=\!
  \frac{16\pi^2 F}{\tau a^2 M^2} \delta\!\left(\!\frac{k}{a}-\frac{M}{2}\!\right)
\end{equation}
and can be directly integrated for each $k$ individually.  The
$\delta$-function in (\ref{regeq}) gives the moment, $\eta=\eta_*$,
when the particle with properly rescaled 3-momentum
\begin{equation}
  \label{matching}
  p=\frac{k}{a}=\frac{a_*}{a}\,\frac{k}{a_*}=\frac{M}{2}\,\frac{a_*}{a}
\end{equation}
was created, so the integration leads to
\begin{equation}
  f_\DM(k) = \frac{16\pi^2F(\eta_*)}{\tau a_*^2 M^2}. 
\end{equation}
The conformal time (for some given number of d.o.f.\ encoded in $c_*$)
is $\eta_*=a_*/c_*$ and \eqref{matching} implies $\eta_*=2k/(c_*M)$.
Then for the Hubble parameter one obtains from \eqref{Hubble}
\begin{equation}
  H_* = \frac{1}{c_*\eta_*^2} = \frac{c_*}{a_*^2}
  = \frac{c_* M^2}{4k^2}
  = \frac{c_*}{c} \frac{H M^2}{4p^2}.
\end{equation}
Using (\ref{Fnrdecay}) we get
\begin{multline}
  \label{fDMNr}
  f_{\text{DM}}(k) = \frac{32\pi^2\widetilde F_i}{\tau M^3}
  \frac{1}{a_*^3H_*}\times 
  \e^{-\frac{1}{2\tau H_*}}
  \\
  = \frac{1}{a^3}\frac{16\pi^2}{\tau
  M^2}\,\widetilde F_i 
  \frac{c}{c_*H}\frac{1}{p}\times
  \e^{-\frac{c}{c_*}\frac{2p^2}{\tau HM^2}}.
\end{multline}
Finally, the DM distribution in physical momentum $p=k/a$ (introducing
the effective number of degrees of freedom in the plasma $g_*(T)$,
photon temperature $T$ and
$M_{\text{Pl}}^*(T)\equiv M_{\text{Pl}}/1.66\sqrt{g_*(T)}$, so that
$H=T^2/M_{\text{Pl}}^*$) reads
\begin{multline}
  \label{fDMNrTT}
  f_{\text{DM}}(p)  = \frac{16\pi^2}{\tau
    M^2}\,\widetilde F_i \l\frac{g_*(T_*)}{g_*(T)}\r^{2/3}
  \frac{M^*_{\text{Pl}}}{T^3}
  \\
  \times \frac{T}{p}\times
  \e^{-\l\frac{g_*(T_*)}{g_*(T)}\r^{2/3}\frac{p^2}{T^2}
    \frac{2M^*_{\mathrm{Pl}}}{\tau M^2}},
\end{multline}
which precisely coincides with the results from
\cite{Kaplinghat:2005sy}.

\section{Applicability of the non-relativistic approximation}
\label{sec:concl}

\begin{figure}[t]
  \includegraphics[width=\columnwidth]{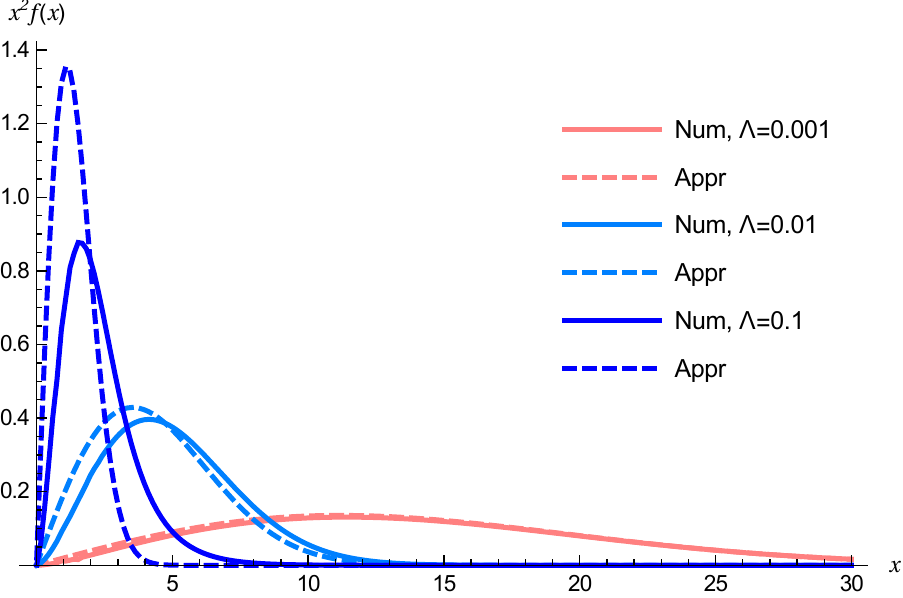}
  \caption{DM spectra (arbitrary units) obtained by exploiting the
    approximate formula (\ref{fDMNrTT}) (dashed curves) and by
    integrating the equations numerically (solid curves), through
    inserting \eqref{sol-2} into \eqref{DMeq}.  Parameter $\Lambda$
    (\ref{appl}) takes three different values: the smaller the value
    is, the more shallow the curves are. The horizontal axis is
    $x=(g_*(T_*)/g_*(T))^{1/3}p/T$.}
    \label{Fig:spectra}
\end{figure}

The final spectrum (\ref{fDMNrTT}) is valid as far as the
non-relativistic approximation (\ref{NrCi}) for the DDM distribution
was applicable.  This puts two bounds on the allowed values of the
momenta $p$ of the DM particle.  The \emph{upper} bound is rather
trivial and is irrelevant for most considerations, as far as it kicks
in the region where the spectrum is anyway exponentially suppressed.
Above the value
\begin{equation}
  \label{high-p}
  p_{\text{max}} = M/2,\qquad f(p_{\DM}>p_{\text{max}}) = 0.
\end{equation}
the result is completely cut-off due to absence (or exponential
suppression in precise calculation) of high momenta DDM particles at
present time.  The more interesting bound modifies the spectrum for
the \emph{low} values of momenta\footnote{We do not discuss here the
  issue of Pauli blocking at low momenta which might be relevant for
  any mechanism with light fermionic dark matter.}.  This is not a
hard cut-off, but a suppression of the spectrum.  This suppression is
not grasped by (\ref{fDMNrTT}) as it is obtained in the assumption
that at the moment $\eta_*$ the DDM particle was already
non-relativistic, and hence the DM momentum at the moment of
production is
\begin{equation}
  p_{\DM}|_{\eta=\eta_*} = \frac{M}{2} \gg \vev{p_\text{DDM}}\big|_{\eta=\eta_*}.
\end{equation}
Later the Universe expands and the DM momentum gets redshifted, so
that at temperature $T$ (today) it equals
\[
  p_{\DM}=\l\frac{g_*(T)}{g_*(T_*)}\r^{1/3}\frac{T}{T_*}\frac{M}{2},
\]
Smaller momenta correspond thus to higher Universe temperature $T_*$
at the time of production.  As far as the typical momentum of a
particle in the plasma is $p\sim 3T$, the DDM particles are
non-relativistic only at $T_*\ll M/3$, and the spectrum
(\ref{fDMNrTT}) is valid only for high momenta,
\begin{equation}
  \label{low-p}
  p_{\DM} \gg \frac{3}{2}\l\frac{g_*(T)}{g_*(T_*)}\r^{1/3} T .
\end{equation}
So the cold part of the distribution is not described by
(\ref{fDMNrTT}), and analysis beyond the non-relativistic
approximation (\ref{NrCi}) is required.  This constraint alone is also
not always an important correction for the spectrum (\ref{fDMNrTT}),
as far as the spectrum is again suppressed at low momenta, at least
for large $\tau$, see e.g.\ Fig.~\ref{Fig:spectra} and the average
momentum \eqref{app-velocity}.

There are two more constraints referring to the model parameters,
which leave the spectrum (\ref{fDMNrTT}) valid.  One is very relevant
in practice.  It follows from the analysis of the assumption of
decaying particle being (almost) at rest and implies the lower bound
on the DDM lifetime,
\begin{equation}
  \label{appl}
  1\ll\tau H(T=M/3)=\frac{2\tau M^2}{18M_{\text{Pl}}^*(T=M/3)}\equiv
  \frac{1}{18} \frac{1}{\Lambda},
\end{equation}
which limits significantly the exponential factor $\propto \Lambda$ in
(\ref{fDMNrTT}).  Note, that the limit corresponds to explosion of the
expansion in the pre-exponent in (\ref{non-rel-correction}).
Basically, the DDM particles with short lifetime decay mostly before
becoming non-relativistic.  So spectrum (\ref{fDMNrTT}) is justified
only for $\Lambda\ll 0.05$, otherwise we come out of the applicability
region.

On the contrary, for very long lifetime $\tau$ of the DDM particle
(small $\Lambda$) and large enough initial abundance the DDM may start
to dominate the Unverse expansion and lead to a temporary matter
dominated stage.  This means that for large $\tau$ formula
(\ref{fDMNrTT}), which is derived assuming radiation domination stage,
also can not be applied.

We illustrate the statements above with Fig.~\ref{Fig:spectra}. 
Here, for several values of $\Lambda$ we plot DM spectra
(\ref{fDMNrTT}) and the exact numerical solution of equation
\eqref{DMeq}, with DDM spectra given by \eqref{sol-2} (see
\cite{Petraki:2007gq}).  The maximum of the distribution moves towards
larger momenta when $\Lambda$ diminishes.  The smaller the latter is
the more accurate the approximation \eqref{appl} becomes.  For larger
values of $\Lambda$ pronounced deviation develops between the
approximate formula (\ref{fDMNrTT}) and the exact numerical spectrum.
At low momenta, $p/T\lesssim 3/2(g_*(T)/g_*(T_*))^{1/3}$ the
approximate spectrum (\ref{fDMNrTT}) is always incorrect, since this
region is beyond the border \eqref{low-p}.

\begin{figure}[!t]
  \includegraphics[width=\columnwidth]{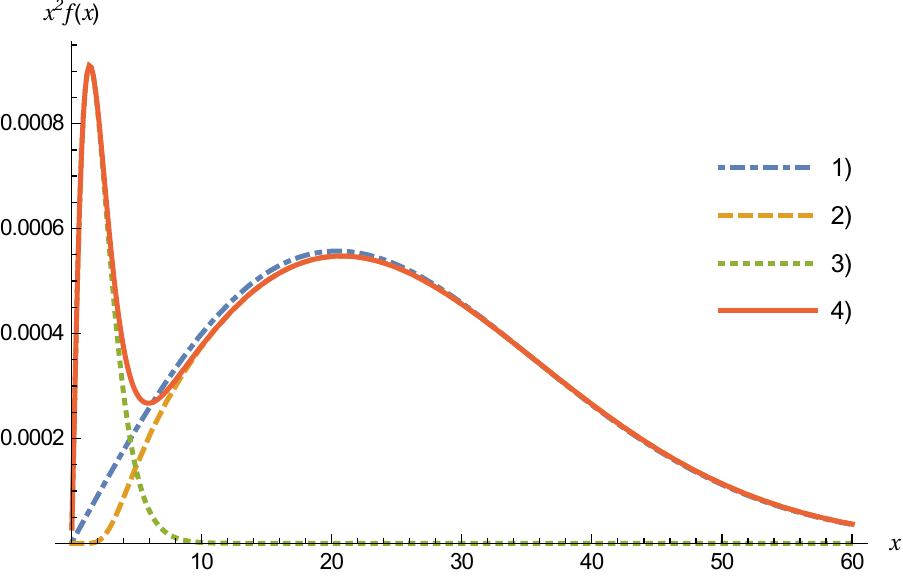}
  \caption{DM spectra obtained (the same variable $x$ as in
    Fig.\,\ref{Fig:spectra}): 1) by exploiting the approximate formula
    (\ref{fDMNrTT}); 2) by integrating the equations numerically,
    through inserting \eqref{sol-2} into \eqref{DMeq}; 3) from decays
    of a particle while it is still in thermal equilibrium (here it
    was assumed to freeze-out at $T\sim M/7.5$), see
    e.g.\,\cite{Shaposhnikov:2006xi}; 4) by summing contributions (2)
    and (3), total answer, see e.g.~\cite{Petraki:2007gq}.  We choose
    $\Lambda=3\times10^{-4}$, which obeys \eqref{appl}.}
  \label{Fig:spectra2}
\end{figure}

Account for these limits impacts the calculations of the average
velocity,
\[
  \la v_{\DM}\ra =\frac{\la p_{\DM}\ra}{T}\,\frac{T}{m_{\DM}}.
\]
When integration is performed using distribution \eqref{fDMNr} over
all momenta (i.e.\ violating the bounds
(\ref{high-p}),~(\ref{low-p})), and/or ignoring the constraint
\eqref{appl} (e.g.\
\cite{Kamada:2013sh,Merle:2013wta,Adulpravitchai:2014xna,Merle:2014xpa})
the obtained average velocity is not correct.  Recall, that the
average velocity is usually adopted in estimates of the free streaming
length important for the small scale structure formation and tested
with Lyman-$\alpha$ forest data.  The \emph{coldest} component of the
dark matter may be obtained for decays of DDM which just start to be
non-relativistic.  To grasp this part of the spectrum the exact
solution of (\ref{DMeq}) is required instead of the approximation
(\ref{fDMNrTT}), see \cite{Petraki:2007gq} for detailed solution.
Even more interesting situation happens if the DDM particle stays in
thermal equilibrium long enough, and decays (partially) into the DM
while still in thermal equilibrium with the rest of the Universe, as
analysed in \cite{Kusenko:2006rh,Shaposhnikov:2006xi}.  This situation
corresponds to using the purely thermal DDM distribution instead of
(\ref{sol-2}) for early production times (or, equivalently, restoring
the full collision integral in the r.h.s.\ of (\ref{eqDDM}), including
all production and destruction terms), and the relevant contribution
to the DDM can be read of \cite{Kusenko:2006rh}.  Then, an additional
peak in the spectrum arises at low momenta, leading to an interesting
two-component DDM spectrum with very pronounced two maxima, see
Fig.~\ref{Fig:spectra2}.  This situation was also analysed in detail
numerically in \cite{Merle:2015oja} for the case of DDM decaying only
to DM particles.

\begin{figure}[tb]
  \includegraphics[width=\columnwidth]{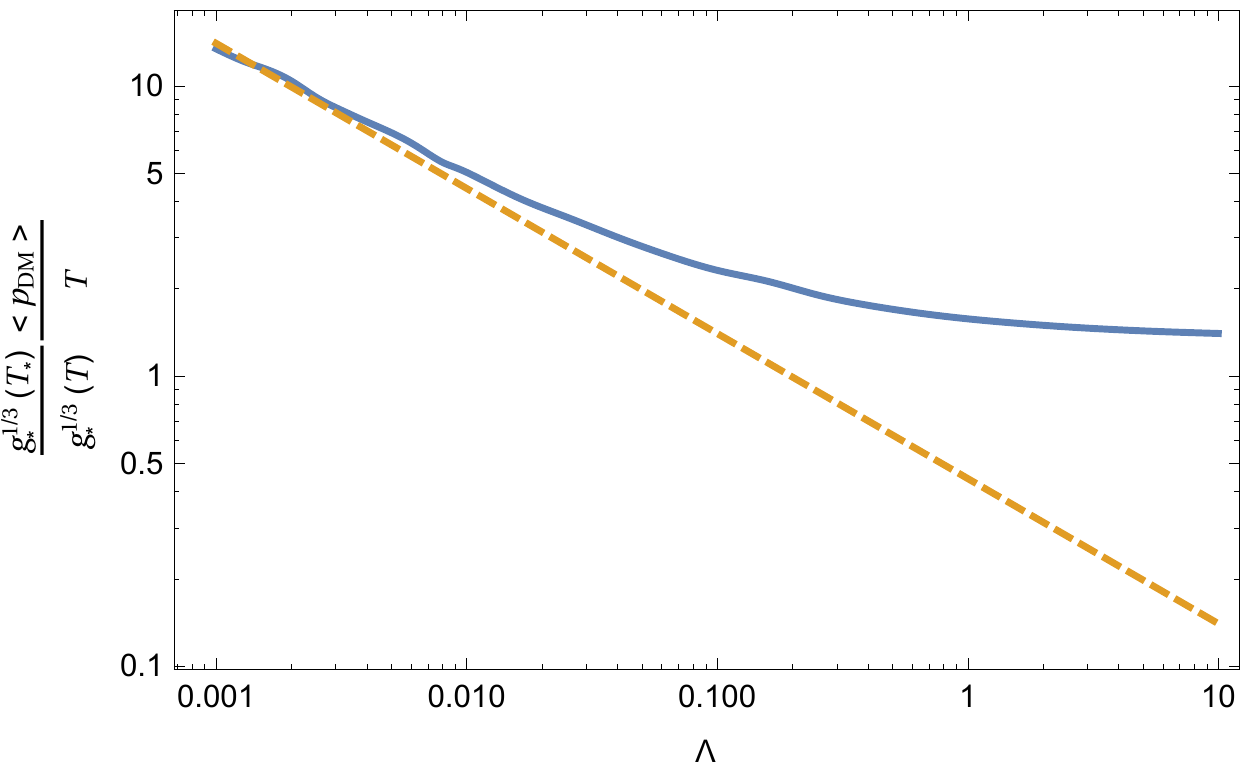}
  \caption{Average velocity of the DM calculated with: formula
    \eqref{app-velocity} derived for the approximated spectrum
    (\ref{fDMNrTT}) (dashed line); 2) numerical integration with
    spectrum obtained through inserting \eqref{sol-2} into
    \eqref{DMeq} (solid curve).}
  \label{Fig:velocity}
\end{figure}

The average velocity for different spectra can differ significantly.
As we show below, it takes place even in the one-component case of
Fig.~\ref{Fig:spectra}.  For the approximate solution (\ref{fDMNrTT})
the average momentum corresponds to
$p/T\simeq (g_*(T)/g_*(T_*))^{1/3}/\sqrt{\Lambda}$ and should always
be much larger than one, if the applicability condition \eqref{appl}
is satisfied.  Therefore, the relative contribution of low momenta
$p/T\lesssim (g_*(T)/g_*(T_*))^{1/3}$ is small, and one can neglect
the bound \eqref{low-p} and obtain the approximation for the average
momentum using the distribution (\ref{fDMNrTT}):
\begin{equation}
  \label{app-velocity}
  \frac{\la p_{\DM}\ra}{T}=\frac{\sqrt{\pi}}{2\sqrt{\Lambda}} 
  \l\!\frac{g_*(T)}{g_*(T_*)}\!\!\r^{\!\!1/3} \!\!\!=\! 
  \frac{\sqrt{\pi\,\tau}M}{\sqrt{2\,M_{\text{Pl}}^*}}\l\!\frac{g_*(T)}
  {g_*(T_*)}\!\!\r^{1/3}\!\!,  
\end{equation}
where presently $g_*(T=T_0)=43/11\approx3.9$ (and not 3.36 as in
\cite{Petraki:2007gq,Kamada:2013sh,Merle:2013wta,Adulpravitchai:2014xna,Merle:2014xpa}).
This value must be compared to the exact numerical estimate obtained
by averaging over the spectra derived by inserting \eqref{sol-2} into
\eqref{DMeq}.  The both results depend on the parameter $\Lambda$, and
coincide when it is sufficiently small, $\Lambda\ll0.05$, so that the
approximation \eqref{appl} is valid.  When $\Lambda$ increases, the
numerical estimate reaches the finite asymptote
$\langle p_{\DM}\rangle/T\sim1.3\times(g_*(T)/g_*(T_*))^{1/3}$, while
the approximate formula (\ref{fDMNrTT}) yields steadily decreasing to
zero average velocity \eqref{app-velocity}, which would imply colder
and colder DM, see Fig.~\ref{Fig:velocity}.
Therefore, the decay of the DDM particles after they leave thermal
equilibrium can lead to DM slightly colder, than a thermally produced
one with $\langle p\rangle/T\sim 3.15\times (g_*(T)/g_*(T_*))^{1/3}$. 

To summarise we conclude that the process of non-thermal generation of
the DM in decays of another particle has many non-trivial features, and
should be approached with care.  The situations that can be tracked
analytically are the case of in-equilibrium decay, leading to the
colder component of the DM, and decay of relatively long-lived
particles that decay out-of-equilibrium when they became
non-relativistic, leading to a relatively hot DM component.
Intermediate situations of short DDM lifetime should be analysed
exactly numerically, as in \cite{Kawasaki:1992kg,Petraki:2007gq,Merle:2015oja}.

\paragraph*{Acknowledgements}
The work of D.G. was supported by the RSCF grant 14-12-01430.

\bibliographystyle{apsrev4-1}
\bibliography{local}

\end{document}